\documentstyle[psfig,epsf]{lamuphys}
\begin{document}
\title{FERMIONIC QUANTUM SPIN GLASS TRANSITIONS}
\author{Reinhold Oppermann and Bernd Rosenow}
\institute{Institut f\"ur Theoretische Physik,
Universit\"at W\"urzburg, D--97074 W\"urzburg}
\maketitle
\begin{abstract}
This article reviews recent progress of the analytical theory of quantum spin
glasses (QSG). Exact results for infinite range and one loop
renormalisation group calculations for finite range models of either
insulating or metallic type are presented.
We describe characteristics of fermionic spin glass
transitions and of fermionic correlations which are affected by these
transitions and by spin glass order. Connections between tricritical
thermal-- and $T=0$ QSG transitions are described.
A general phase diagram with tricritical QSG 
transitions caused either by random chemical potential or 
by elastic electron scattering, 
and implying discontinuous $T=0$--transitions in weak and in strong filling 
regimes, is also derived.

\end{abstract}

\section{Introduction}
Fermionic quantum spin glasses form a part of the overlap regime of
interacting disordered fermion systems and spin glasses. 
Understanding such systems with frustrated or with general random interactions 
between fermionic spins touches many currently active research fields.
The wellknown transverse field Ising spin glass can also be represented 
as a pseudofermionic quantum spin glass with imaginary chemical potential and
thus be treated analogously to the ones discussed here which involve real
fermion spins.
One of the attractive features of fermionic spin glasses is the connection of
spin and charge degrees of freedom. Even in a strongly localized Ising limit
(with all spin and charge variables commuting) quantum
statistics remains indispensable, since it governs the relative occupation of
magnetic and nonmagnetic states. This provides a link between
spin-- and charge correlations, exerting a
major influence on magnetic phase diagrams for example. 
Nonanalytic interplay 
is observed in the
simplest fermionic spin glass, the SK--model
extended to a four--state--per--site fermionic space ($ISG_f$). This phenomenon occurs at a
thermal tricritical point, separating continuous spin
glass transitions around half--filling from discontinuous
regimes at either strong or weak filling. This structure of the phase
diagram is compared with a similar one of classical spin 1 models including
the clean BEG--model. 
Furthermore, model definitions on fermionic space generate 
a large set of correlation functions 
which display ubiquitous quantum dynamics 
of these models on the fermionic level, 
ie beyond the one seen in spin-- and charge--correlations and 
caused by electronic transport.
The fermionic Green's function and its density of states content 
is one of those quantities which probe the commutation
properties of single fermion operators with the hamiltonian. 
The $ISG_f$ shows such quantum--dynamics.
Moreover, correlations of the statistical density of states fluctuations 
turn out to be related to the spin glass order parameter, 
which introduces Parisi replica symmetry breaking into the charge sector. 
\\
\section{The Grand Canonical Fermionic Ising Spin Glass}
In contrast to standard spin glass models like SK-- or transverse field
models, the models of interacting true fermion spins are naturally described
in the grand canonical ensemble. Statistically fluctuating spin interactions
leave only the choice between nonrandom local fermion concentration or
nonrandom chemical potential. 
In any case however an effective spin dilution
due to thermal redistribution among magnetic and nonmagnetic states is
controlled by either $\mu$ or $\nu$. This form of thermal spin dilution
comprises at a special point the case of nonanalytically communicating
charge-- and spin--fluctuations. The simplest
model displaying this type of behavior is the 
Ising spin lass on a fermionic space ($ISG_f$) with four states per site,
defined by
the hamiltonian
$H = -\frac{1}{2} \sum_{i \neq j} J_{i j} \sigma_i \sigma_j - \mu \sum_i
n_i$ with
spins $\sigma_i =  \Psi^{\dagger}_{i, \alpha} \sigma^z_{\alpha \beta}
\Psi_{i, \beta}$ , particle number operator $n_i= \Psi^{\dagger}_{i, \alpha}
\Psi_{i, \alpha}$  and gaussian distributed exchange integrals
$J_{i j}$ \cite{opgro}. 
The fermionic field operators obey the usual commutator
relations $ \{ \Psi_{i \alpha}, \Psi_{j \beta} \} =0$ and
$ \{ \Psi^{\dagger}_{i \alpha}, \Psi_{j \beta} \} = \delta_{i j} \delta_
{\alpha \beta} $.
Expressing the partition function with the help of Grassman integrals
the disorder averaging of the free energy is performed by means of the
replica trick $\beta [F]_{av}= [ \log Z]_{av}=\lim_{n \to 0}\frac{1}{n}
(1-[\Pi_{\alpha =1}^n
Z^{\alpha}]_{av})$.
As a guide to the global phase diagram we study an exactly solvable
infinite range version
of the model but the formulae obtained in this subsection may equally well
be considered as saddle point approximation for an interaction with finite
range.
\subsection{Phase Diagram}
Details of the calculation are given in \cite{opgro}, here
we just state the result for the replica symmetry broken saddle point free 
energy  
%------------------------equation 1 ------------------------------------------
\begin{eqnarray}
\beta f & = & \frac{1}{4} \beta^2 J^2 \left[(1-\tilde{q})^2 -(1-q_1)^2 +
q_1^2-\int_0^1  dx\, q^2(x)\right]
-\ln 2 - \beta \mu\nonumber\\
&  &- \lim_{K \to \infty} \int_{z_{K+1}}^G \ln \left[ \int_{z_K}^G
\left[...\left[ \int_{z_1}^G \left( \cosh(\beta \tilde{H}) 
\right.\right.\right.\right. \label{one}\\
&  & \left. \left. \left. \left.
+ \cosh(\beta \mu)\exp(-\frac{1}{2}\beta^2 J^2(\tilde{q}-q_1)) \right)^{
m_1} \right]^{m_2/m_1} ... \right]^{m_k / m_{K-1}} \right]^{1/ 
m_K}\nonumber
\end{eqnarray}
%----------------------------------------------------------------------------
Notice the appearance of the Parisi variables $q_\nu$ and the additional
$\tilde{q}$ which lies at the heart of the following discussion. The 
Edwards - Anderson order parameter $q_{EA} = \lim_{t \to \infty} <S_i(t)
S_i(0)>$ describes the freezing of spins in the spin glass phase and is 
given by the Plateau height
$q(1)$ of the order parameter function, 
whereas the replica diagonal $\tilde{q} = [< \sigma^a 
\sigma^a>]_{av}$ 
is related to the average filling factor $[\nu]_{av} = \frac{1}{N} \sum_i
[n_i]_{av }$ by 
$[\nu]_{av} =1+ \tanh(\beta \mu)(1 - \tilde{q})$. The last relation is exact 
even in the case of replica symmetry breaking. To obtain information about
the phase diagram a replica symmetric approximation is sufficient, though.
The symmetric saddle--point solutions are 
%--------------------------equations 3,4,5-----------------------------------
$q = \int_z^G  \frac{\sinh^2[\beta \tilde{H}(z)]}{{\cal C}^{2}_{\mu}(z)}$
and $\tilde{q} = \int_z^G \frac{ \cosh[\beta\tilde{H}(z)]}
{{\cal C}_{\mu}(z)} $
%----------------------------------------------------------------------------
with ${\cal C}_{\mu} (z) =  \cosh[\beta\tilde{H}(z)] +  
\cosh(\beta\mu)\exp[-1/2\beta^2(\tilde{q}- q)]$. Phase transitions 
are signalized by vanishing masses of the order parameter
propagators which in the saddle point formalism are given by second 
derivatives of the free energy. 
Hence a positive mass for $\tilde{q}$ and a 
negative one for $q$ guarantees stability. 
A similar system of coupled stability conditions was found for the 
BEG - model \cite{BEG} and for a SK - model with crystal 
field \cite{MoShe}. 
Analyzing the stability 
limits
one obtains curves of 
critical spin and charge fluctuations, respectively:

%*************************equation 10***************************************
\begin{eqnarray}
\mu_{c1}(T)=T \cosh^{-1}[(1/T-1)\exp[1/(2T)]]
\label{ten}
\end{eqnarray}
%******************************************************************************
%****************************equation 11************************************
\begin{eqnarray}
\mu_{c2}(T)=T \cosh^{-1}[\frac{(1\mp\sqrt{1-8T^2})^2}{8T^2}
\exp[\frac{2}{1\mp\sqrt{1-8T^2}}]]
\label{eleven}
\end{eqnarray}
%******************************************************************************
The two curves have a common tangent point at
$\mu_{c3}=1/3 \cosh^{-1}[2\exp(3/2)]=0.961056, T_{c3}=1/3$.
For smaller values of the chemical potential the paramagnetic solution 
 becomes unstable to spin fluctuations first, at the common 
tangent point both types of fluctuations
become critical simultaneously and tricritical behaviour results. 
\subsection{Tricritical Point TCP: Exponents and Special Features}% near the TCP} 
An expansion of the saddle point equations around the 
tricritical point 
%at $T_{c3}=J/3, \mu_{c3} = \frac{J}{3} \cosh^{-1}(
%2 \exp(\frac{3}{2})) \simeq .9611J$ 
yields in leading order 
%*****************************equations 19, 20*****************************
\begin{eqnarray}
0=g r_g - r_T \delta T^2+6 \delta T \delta\tilde{q}- 
\frac{3}{4}\delta\tilde{q}^2+ 3 q^2
\label{nineteen},\quad\quad
0=6 q (\delta\tilde{q}-\delta T-q)%\label{twenty}
\end{eqnarray}
%***************************************************************************
where $\delta\tilde{q}\equiv\tilde{q}-\tilde{q}_{TCP}, 
gJ=\mu-\mu_{c3}+(\zeta^{-1}J-\mu_{c3})3 \delta T$ as nonordering field, and 
$\delta T \equiv T-T_{c3}$. The constants are given by 
$r_g=\frac{2\zeta}{3}, 
r_{T}=2(1-\frac{3}{4}\zeta^{-2})$ 
with $\zeta\equiv tanh(\mu_{c3}/T_{c3})\simeq 0.9938$.
The average filling factor corresponding to $\mu_{c3}$ is evaluated
as $[\nu_{c3}]_{av}\simeq 1.6625$. From (\ref{nineteen}) we get for q=0
%**************************equation 21 paramagnetic solution**************
\begin{eqnarray}
\delta\tilde{q}_{dis}=4(\delta T \pm|\delta T|\sqrt{1+\frac{r_g g}
{12 \delta T^2}-\frac{r_T}{12}}=4(\delta T \pm |\delta T| W)
\label{twentyone}
\end{eqnarray}
%**************************************************************************
Only the solution  with the - sign corresponds to a minimum
of the free energy,
in a region close to the 
line $\delta \mu = - 3(1/\zeta - \mu_{c3}) \delta T= - 0.1354 \delta 
T$  (tangent to both $\mu_{c1}(T_{c3})$ and $\mu_{c2}(T_{c3}))$ g is of
order $\delta T^2$ or smaller and usual critical behaviour results.
However, if g is of order $\delta T$ or larger the solution becomes to
leading order $\delta \tilde{q} = \sqrt{\frac{r_g g}{12}}$ and thus displays 
a nonanalytical dependence on temperature and / or chemical potential. This 
type of crossover can also be seen from the scaling form of the free energy 
$f_{dis}=|\delta T|^{2-\alpha} {\cal G}(\frac{g}{\delta T^2})$
which allows for the identification of the specific heat exponent
$\alpha =-1$ and the crossover exponent $\phi = 2$. The crossover function
${\cal G}(x)$ is regular for small x and  has the asymptotic form
${\cal G}(x) {\approx  \atop x\to\infty} (\frac{g}{\delta T^2})^{\frac{2-\alpha}
{\phi}} ({\cal G}_{\infty}+$regular corrections$)$. In the tricritical region the
leading singularity in the free energy is given by
$f_{TCP, sing}=\frac{4}{\sqrt{3}}(r_g g)^{\frac{3}{2}}$.
For $\delta\mu=0$ we have $g\sim \delta T$ and can read off the tricritical
specific heat exponent $\alpha_3=\frac{1}{2}$ from above.

Using $q=\delta\tilde{q}-\delta T$ one finds ordered solutions displaying
the same crossover as discussed above, the transition from paramagnet to spin
glass
being continous for positive $\delta T$ and first order for negative
temperature deviations. 
In the tricritical regime we find
%-----------------------------equation 26--------------------------------
%\begin{eqnarray} 
$q=\delta\tilde{q}=\frac{2}{3}\sqrt{-r_g g}$
%\label{twentysix}
%\end{eqnarray}
%-------------------------------------------------------------------------
%
which yields the tricritical order parameter exponent $\beta_3=\frac{1}{2}$
and suggests that $q$ and $\delta \tilde{q}$ act as order parameters 
simultaneously. From the fluctuation Lagrangian (see section 3) one reads 
off mass squared
proportional to $\delta T$ and hence $\gamma_3 = \beta_3 = \alpha_3 = 
\frac{1}{2}$.

\section{Tricritical Landau Theory and the Parisi Solution of the Fermionic 
Ising Spin
Glass}
We derived a fluctuation theory for the tricritical and finite
range $ISG_f$; 
for $T\neq 0$ 
a Lagrangian of the same structure is obtained for
models including transport mechanism by integrating
out dynamical degrees of freedom
\begin{eqnarray}
L&=& \frac{1}{t}\int d^dx{\large[}
-\frac{3 h^2}{2 J} \sum Q^{ab} +\frac{r\kappa_1}{(\kappa_2)^2}\sum Q^{aa}   
+ \frac{1}{2}\sum Q^{aa}(-\nabla^2+u)Q^{aa}
\nonumber\\ 
&+&\frac{1}{2}Tr^{\prime}(\nabla Q^{ab})^2
- \frac{1}{t}\sum^{\hspace{.6cm}\prime} Q^{aa}Q^{bb}
-\frac{\kappa_1}{3}\sum (Q^{aa})^3\nonumber\\
&-&\frac{\kappa_3}{3}
Tr^{\prime}Q^3 
-\kappa_2\sum^{\hspace{.5cm}\prime}
Q^{aa}Q^{ab}Q^{ba}+\frac{y_4}{4}\sum^{\hspace{.5cm}\prime} (Q^{ab})^4
{\large],}
\label{eighteen}
\end{eqnarray}
Here
$4(\frac{\kappa_1}{t})^{(0)}=(\frac{\kappa_2}{t})^{(0)}=
(\frac{\kappa_3}{t})^{(0)}
=\frac{3^3}{2}$
and $u^{(0)}=0$ denote the bare coefficients at tricriticality.
One fourth order term relevant for replica symmetry breaking is kept.
Replicas under
$\sum^{\prime}$ or $Tr^{\prime}$ are distinct. The $Q^{aa}Q^{bb}$--coupling
is renormalization group generated as in the metallic quantum spin
glass, its effects will be discussed in a subsequent section.

\unitlength1cm
\hspace{-.8cm}
\begin{minipage}[t]{6cm}
\psfig{file=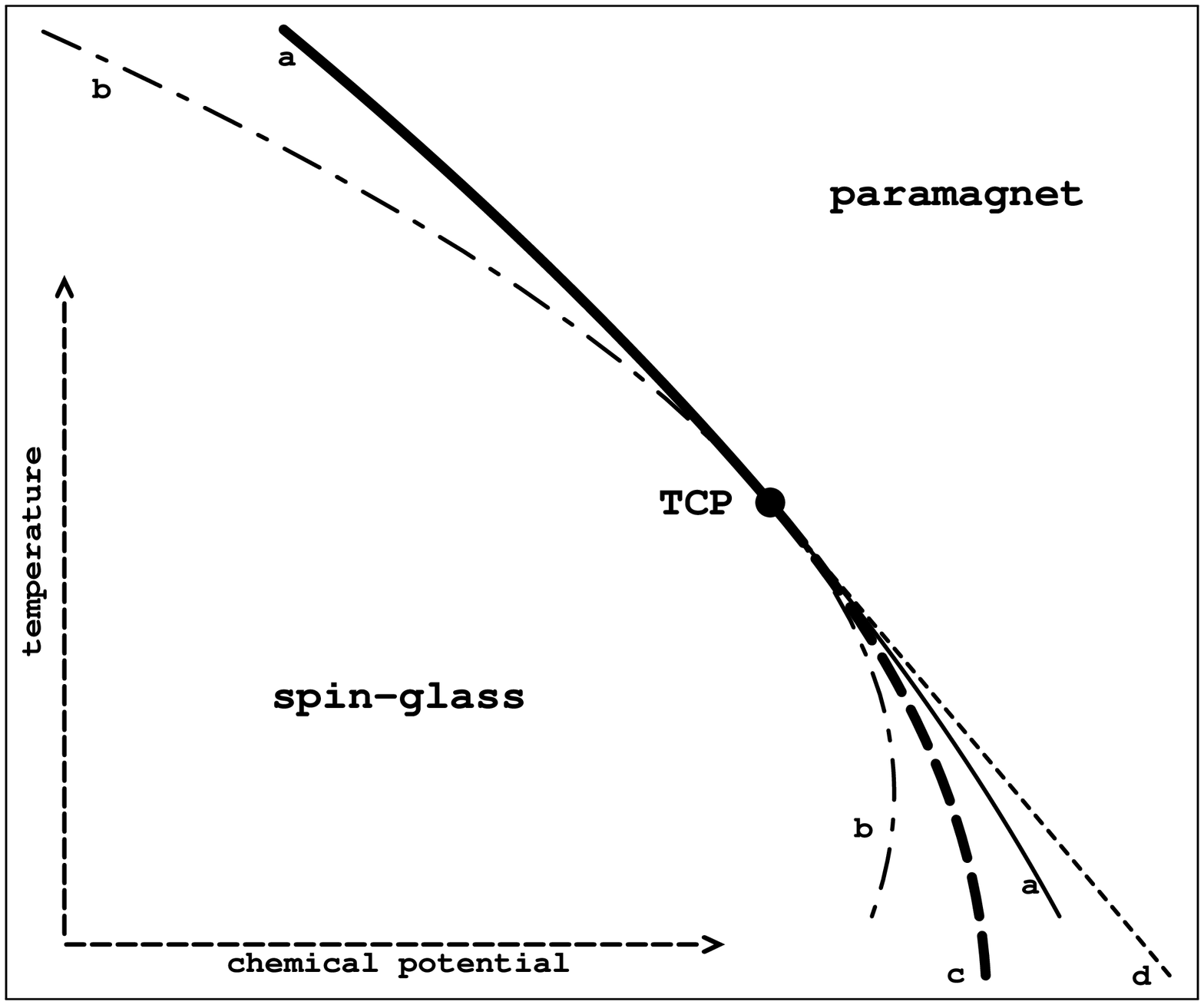,width=6cm,angle=90}
\end{minipage}
\vspace{-7.2cm}
\begin{figure}
\hfill
\begin{minipage}[t]{3.6cm}
\caption{\it Vicinity of the tricritical point ($TCP$) for positive chemical
potential.
Continuous spin glass transitions occur on curve (a) above the $TCP$
(thick unbroken line).
Below the $TCP$ first order thermodynamic transitions take place
on curve (c). Curve (d), starting at the $TCP$, and curve (b) below the $TCP$
limit the existence regime of ordered and disordered phases respectively.}
\end{minipage}
\end{figure}

Inserting the Parisi Ansatz for replica symmetrie breaking in the Lagrangian
(\ref{eighteen}) the  
expansion of the replica symmetry broken saddle point free energy 
around the tricritical point reads
%***************************equation 17***********************************
\begin{eqnarray}
f&-&f_{TCP}=\mu -\mu_{c3}-\frac{3 h^2}{2 J} [\delta\tilde{q}-\int_0^1 q(x)
dx] + J\{(\frac{3}{2}r_g g-r_{\tau}\tau^2)\delta\tilde{q}\nonumber\\
&+&\frac{3}{2}\tau[(\delta\tilde{q})^2 -\int^{1}_{0}dxq^2(x)]
-\frac{3}{2}
[\int^{1}_{0}dx[xq^3(x)+3q(x)\int^{x}_{0}dyq^2(y)]\nonumber\\
&-&3\delta\tilde{q}\int^{1}_{0}dxq^2(x)
+\frac{1}{4}(\delta\tilde{q})^3]-\frac{y_4}{4}\int^{1}_{0}dxq^4(x)\} ,
\label{seventeen} 
\end{eqnarray}
%*****************************************************************************
Here variables and constants are defined as in (\ref{eleven}), $\tau =\delta
T / T_{c3}$.
In contrast to crystal--field split spin glasses
\cite{MoShe} the quartic coefficient $y_4$ of our free
energy,
Eq.(\ref{seventeen}), is nonzero and one obtains the Parisi solution $q(x)=\frac{9}{2y_4}x$ for 
$0\leq x\leq x_1$ and $q(x)=q(1)$ for $x_1\leq x\leq 1$. The plateau 
height is found to satisfy $q(1)=\delta\tilde{q}+O(\delta\tilde{q}^2)$. 
Consequently, plateau and breakpoint scale like $\sqrt{|\tau|}+O(\tau)$ 
at the $TCP$, while linear $\tau$-dependence is reserved 
to $T_c>T_{c3}$. Adapting the notation of 
\cite{FiSo} we express our result for the irreversible response 
$q(1)-\int^{1}_{0}q(x)\sim |\tau|^{\beta_{\Delta}}$ in terms of the 
exponent $\beta_{\Delta 3}=1$ for $T\rightarrow T_{c3}$ and 
$\beta_{\Delta}=2$ for $T\rightarrow T_c>T_{c3}$.
For the Almeida--Thouless line at tricriticality we find
$\frac{H^2}{J^2}=\frac{80}{81}(\frac{2}{3}(1-\frac{\mu_{c3}}{J}
tanh(\frac{3\mu_{c3}}{J})))^{3/2}\tau_{AT}^{3/2}+O(\tau_{AT}^2)$
with $\tau_{{\small AT}}\equiv \frac{T_{c3}-T_{AT}(H)}{T_{c3}}$. Hence we 
obtain the critical exponent $\theta_{3}=\frac{4}{3}$ near $T_{c3}$, while 
$\theta=\frac{2}{3}$ for all $T_c>T_{c3}$. 
These values do not 
satisfy the scaling relation $\theta_{3}=
\frac{2}{\beta_{\Delta_3}}$ with 
$\beta_{\Delta_3}=1+(\gamma_3-\alpha_3)/2$. Along the lines described in
\cite{FiSo}, this problem of mean--field exponents will be 
resolved 
below by the renormalization group analysis of the coupling $y_4$ of the 
{\it finite--range and finite--dimensional $ISG_f$}.

\section{Replica Symmetry Breaking for Fermions}
The fermionic Ising spin glass allows for an exact evaluation 
of the bare fermion propagator
$G_{ij,\sigma}=<<\frac{1}{i\epsilon_n + \mu + \tilde{H}}>>$
at least in the disordered phase; 
here $\tilde{H}$ denotes the usual effective field and the 
double average refers to the replica--local and the 
Parisi block decoupling fields. The result can be written in the form 
\begin{eqnarray}
G_{ij,\sigma}(\epsilon_n)&=&-i\sqrt{\frac{\pi}{2J\tilde{q}}}
\sum_{\lambda =0,\pm 1} \frac{(2-\lambda^2)e^{\frac{1}{2}(\beta 
J\lambda)^2\tilde{q}}ch((1-\lambda^2)\beta\mu_{\sigma})}{exp(\frac{1}{2}\beta^2
J^2\tilde{q})+ch(\beta\mu_{\sigma})}\\                      
& &(1-erfc(\frac{\epsilon_n-i\mu_{\sigma}+i\lambda\beta 
J^2 \tilde{q}}{J\sqrt{2\tilde{q}}J})) 
exp(\frac{\epsilon_n-i\mu_{\sigma}+i\lambda\beta 
J^2\tilde{q}}{\sqrt{2\tilde{q}}J})^2 \delta_{ij},\nonumber
\end{eqnarray} 
where $\mu_{\sigma}\cong \mu +\sigma H$ includes a magnetic field $H$ and $\epsilon_n=(2n+1)\pi 
k_B T/\hbar$.
One 
easily extracts the disorder averaged electronic density of states \cite{BRRO}
\begin{eqnarray}
<\rho(\epsilon)>=\frac{1}{\sqrt{2\pi\tilde{q}}J}e^{-\frac{(\epsilon+\mu)^2}
{2J^2\tilde{q}}}
\frac{ch(\beta\mu)+ch(\beta(\epsilon+\mu))}{ch(\beta\mu)+ch(\beta 
H)e^{\frac{1}{2}\beta^2 J^2\tilde{q}}}
\end{eqnarray}
Below the freezing temperature $O(q^2)$--corrections  occur in $<\rho>$. Statistical 
fluctuations $\delta\rho=\rho - <\rho>$ of the density of 
states are for example observable in 
$<\delta\rho^a_{\sigma}(\epsilon)\delta\rho^{a^{\prime}}_{\sigma^{\prime}}
(\epsilon^{\prime})>$.
Taking Parisi symmetry breaking into account, this correlation becomes a function of 
the Parisi parameter x.
Picking the one with $\sigma=\sigma^{\prime}$ and $\epsilon=\epsilon^{\prime}$ we obtain the 
(zero field) result
\begin{eqnarray}
<\delta\rho_{\sigma}(\epsilon)\delta\rho_{\sigma}
(\epsilon)>(x)&=&\frac{1}{4\pi J^2\tilde{q}_c^5}
(1 - \tilde{q}_c + \tilde{q}_c e^{-\frac{1}{2\tilde{q}_c}}cosh(\frac{\epsilon + 
\mu}{J\tilde{q}_c}))^2\nonumber \\(1+\tilde{q}_c-\frac{(\epsilon+\mu)^2}{J^2}&)&^2 
e^{-\frac{(\epsilon+\mu)^2}{J^2\tilde{q}_c}}q^2(x) + O(q^3(x)).
\end{eqnarray}
This result shows that the fermionic density fluctuations reflect the irreversible 
magnetic response introduced by the 
spin glass order. It is valid for all fillings, one only has to insert the appropriate Parisi 
function and the 
filling--dependent value of the spin autocorrelation $\tilde{q}_c$ at the critical point; 
this implies that 
the scaling of the density correlator with $T-T_c$ is different in the second order regime 
(quadratic) and at the 
tricritical point (linear), while for the discontinuous regime a one--step RSB is expected. 
It is also clear that for 
models with transport mechanism the calculation of conductance fluctuations is of 
great interest. 
So far we have put aside the question of replica symmetry 
breaking of fermion propagators: 
by this we mean the possibility of a nonvanishing propagator between different 
replicas, ie $<\overline{\psi}^a\psi^{b\neq a}>$. 
In the light of the Mezard--Parisi instability of the random field 
Ising model we feel that this preferably would occur as a fluctuation effect, if at all.

\section{Related Fermionic Spin Models}
\subsection{The Fermionic Ising Chain}
Similarities between the phase diagrams of the clean BEG--model 
and of the fermionic Ising spin glass can be 
taken as indicative for the fact that rather spin dilution than disorder is the 
source of the tricritical crossover from continuous to discontinuous phase 
transitions. 
While it is complicated to solve 1D fermionic Ising spin glasses
exactly, the clean fermionic Ising chain offers some simple exact solutions. 
Here we shall provide insight into the role of 
the chemical potential and moreover generalize known results into the 
complex $\mu$--plane. Lee and Yang
\cite{yanglee} derived the distribution of zeroes of the 
partition function of finite and infinite Ising chains within the complex 
magnetic field plane. 
%In fermionic Ising systems the chemical potential 
%can be seen as complementary to the magnetic field. 
Stimulated by the 
representation of conventional Ising chains by fermionic ones with special 
imaginary chemical potential, one may wish to extend the Yang Lee 
analysis to a fourdimensional space 
of complex $(\mu, \nu)$.
The transfer matrix ${\bf T_f}$ of the fermionic Ising chain 
reads
\begin{eqnarray} T_f = e^{\beta\mu}\left( 
\begin{array}{*{4}{c}} 
e^{\beta\mu} & 1 & 
e^{\frac{1}{2}\beta(\mu +h)} & e^{\frac{1}{2}\beta(\mu-h)} \\
1 & e^{-\beta\mu} & e^{\frac{1}{2}\beta(h-\mu)} & 
e^{-\frac{1}{2}\beta(\mu+h)}\\
e^{\frac{1}{2}\beta(\mu+h)} & e^{\frac{1}{2}\beta(h-\mu)} & 
e^{\beta(J+h)} & e^{-\beta J}\\
e^{\frac{1}{2}\beta(\mu-h)} & e^{-\frac{1}{2}\beta(\mu+h)} & e^{-\beta 
J} & e^{\beta(J-h)}
\end{array} \right) 
\end{eqnarray}
The transfer matrices ${\bf T_f}$ and ${\bf T_s}$ of the standard $S=\pm 1$--chain and 
their eigenvalues do not map onto each other at
$\mu=i\frac{\pi}{2}T$, while the partition functions obey
$Z_f^{(N)}=Tr T_f^N =(2i)^NZ_s^{(N)}(\mu=i\frac{\pi}{2}T)$
for any number N of sites.
The largest eigenvalue determines the free energy of the infinite chain,
while the second largest is required in addition to determine the 
correlation length. 
The eigenvalues for $h=0$ are found as
\begin{eqnarray}
\nonumber
\lambda_{\pm}&=&e^{\beta\mu}[ch(\beta\mu)+ch(\beta
J)\pm\sqrt{(ch(\beta\mu)+ch(\beta J))^2+4 ch(\beta\mu)(1-ch(\beta
J))}]\\
\lambda_0&=&0\quad,\quad  \lambda_1=2e^{\beta\mu}sh(\beta J).
\end{eqnarray}
The correlation length is given by
$\xi = 1/ln(\frac{\lambda_+}{\lambda_1})$.
In the $T\rightarrow 0$--limit a transition arises at $\mu=J$ 
and due to the properties
\begin{eqnarray}
\lambda_1\sim exp(\beta(\mu +J)),\quad \lambda_+\sim 
\left\{ \begin{array}{cc} exp(\beta(\mu+J)),\quad \mu < J\\ 
exp(2\beta\mu), \quad \mu > J \end{array} 
\right.
\end{eqnarray}
Thus $\xi$ diverges only for $\mu<J$,  
since the energy for adding a fermion is larger 
than the gain from a magnetic bond if $\mu>J$. Hence 
\begin{eqnarray}
\xi\sim \left\{ \begin{array}{c} exp(\beta(2J-\mu)) \quad , 0 < \mu < J \\ 
exp(\beta J/2) \quad , \mu = J ,
\end{array} \right.
\end{eqnarray}
while $\xi \sim T/(\mu-J)$ for $\mu>J$.
The filling factor
shows for $T\rightarrow 0$ that the system is completely 
filled for $\mu>J$ (empty for $\mu<-J)$. 
Thus there
is no physical $T=0$--transition of this simple system. The correlation 
length diverges in the $T\rightarrow 0$ limit for all fillings $\nu$.
The zero--field partition function shows that Yang--Lee zeroes approach
$\mu=\pm J$ for $T\rightarrow 0$. This means that for 
$(h=0,T=0)$ nonanalytic behaviour (as a function of the real chemical 
potential) can only occur at the values $\mu =\pm J$.
It is instructive to consider $N=2$ explicitly, which yields
\begin{equation}
Z^{N=2}_f=4e^{2\beta\mu}[(ch(\beta\mu)+ch(\beta J))^2 + ch^2(\beta h)
(e^{2\beta J}-1) - sh(2\beta J)]
\end{equation}
This almost trivial case already shows zeroes at
$\mu_0=\pm(J+(\frac{1}{2}ln2\pm i\frac{\pi}{2})T)$,
while allowing for finite complex magnetic field the first zero different
from $\pm J$ in the $T\rightarrow 0$--limit becomes possible with
$\mu =\pm(i\frac{\pi}{4}+ 2 im\pi)T$.% = \mu + i m\pi T$ .
More zeroes occur on the $T=0$--axis as N is increased. For 
$N\rightarrow\infty$ a density function is expected in accordance with
$\xi$ diverging for any $\mu$.

\subsection{Mapping the twodimensional Ising model with imaginary magnetic
field $h=\frac{i \pi}{2} T$ into fermionic space}
The complementary role of complex magnetic field and complex chemical potential
can nicely be seen by recalling 
the exact solution for the 2d Ising model $h=i\frac{1}{2}\pi T$.
This value corresponds to $\mu=i\frac{1}{2}\pi T$, which maps the fermionic
Ising model onto the one above. Thus the exact solution of the 2d fermionic 
Ising model with $\mu=h=i\frac{1}{2}\pi T$ is known. Moreover this special
model maps onto an interaction model of spinless 
fermions with a special species obeying bare Bose statistics but carrying
along the minus signs of fermion interactions. 
The hamiltonian of this model can be written
\begin{equation}
H=-\sum_{ij}J_{ij}\sigma_i^z\sigma_j^z-\mu\sum_i 
(n_{i\uparrow}+n_{i\downarrow})-h\sum_i \sigma_i^z
\end{equation}
with $\mu=h=i\frac{\pi}{2}T$. 
This reduces to zero the imaginary field of one 
fermionic species, while the other field equals the distance 
between Bose-- and Fermi--Matsubara energies. Setting $c=a_{\uparrow}$ and 
$d=a_{\downarrow}$ the hamiltonian reads
\begin{equation}
H=-\sum_{ij}J_{ij}[c^{\dagger}_i c_i c^{\dagger}_j c_j
- c^{\dagger}_i c_i d^{\dagger}_j d_j - d_i^{\dagger}d_i c_j^{\dagger}c_j 
+
d^{\dagger}_i d_i d^{\dagger}_j d_j] - i\pi T\sum_i 
d^{\dagger}_i d_i.
\end{equation}
The new imaginary chemical potential $-i\pi T$ of d--fermions renders their 
Matsubara energies and single particle statistics bosonic. Thus 
anticommuting d--bosons interact with c--fermions.
All vertices retain the fermionic minus--signs.
In the fermionic path integral the c-- and d--particles are described by
the Grassmann fields 
$\psi_c(\tau)$ and by $\psi_d(\tau)$ respectively. The $i\pi T \hat{n}_d$--term 
can be absorbed by the phase 
transformation $exp(i\pi \tau)\psi_d(\tau)=\tilde{\psi_d}(\tau)$.
The new anticommuting fields $\tilde{\psi_d}$ obey unusual 
{\bf bosonic periodicity} 
$\tilde{\psi}_d(\tau) = \tilde{\psi}_d(\tau +\beta)$ on the imaginary time axis,
while $\psi_c(\tau)=-\psi_c(\tau + \beta)$ remains fermionic.
The bosonic feature of 
the anticommuting fields $\psi_d$'s  
is of course seen in the bare statistics.
Perturbatively the above exact conclusion maps the 2D Ising model
with $h=i\pi T/2$ onto a coupled Fermi--Bose 
system (c--d) with additional 
$(-1)$--factors for each d--loop.

\section{Metallic Quantum Spin Glass}
So far we have discussed the magnetic phase diagram
of fermionic lattice gases having in mind that  the tricritical 
phenomenon and universal quantities are not changed by coherent hopping of
the electrons in flat bands. On the other hand it is well known 
\cite{ROMB}, \cite{SRO} that both a random chemical potential and an 
increasing band width suppress the 
transition temperature
continuously down to zero thereby producing a quantum phase transition (QPT).
The field theory of such a QPT

\vspace{-2.2cm}
\unitlength1cm
\begin{minipage}[]{6cm}
\psfig{file=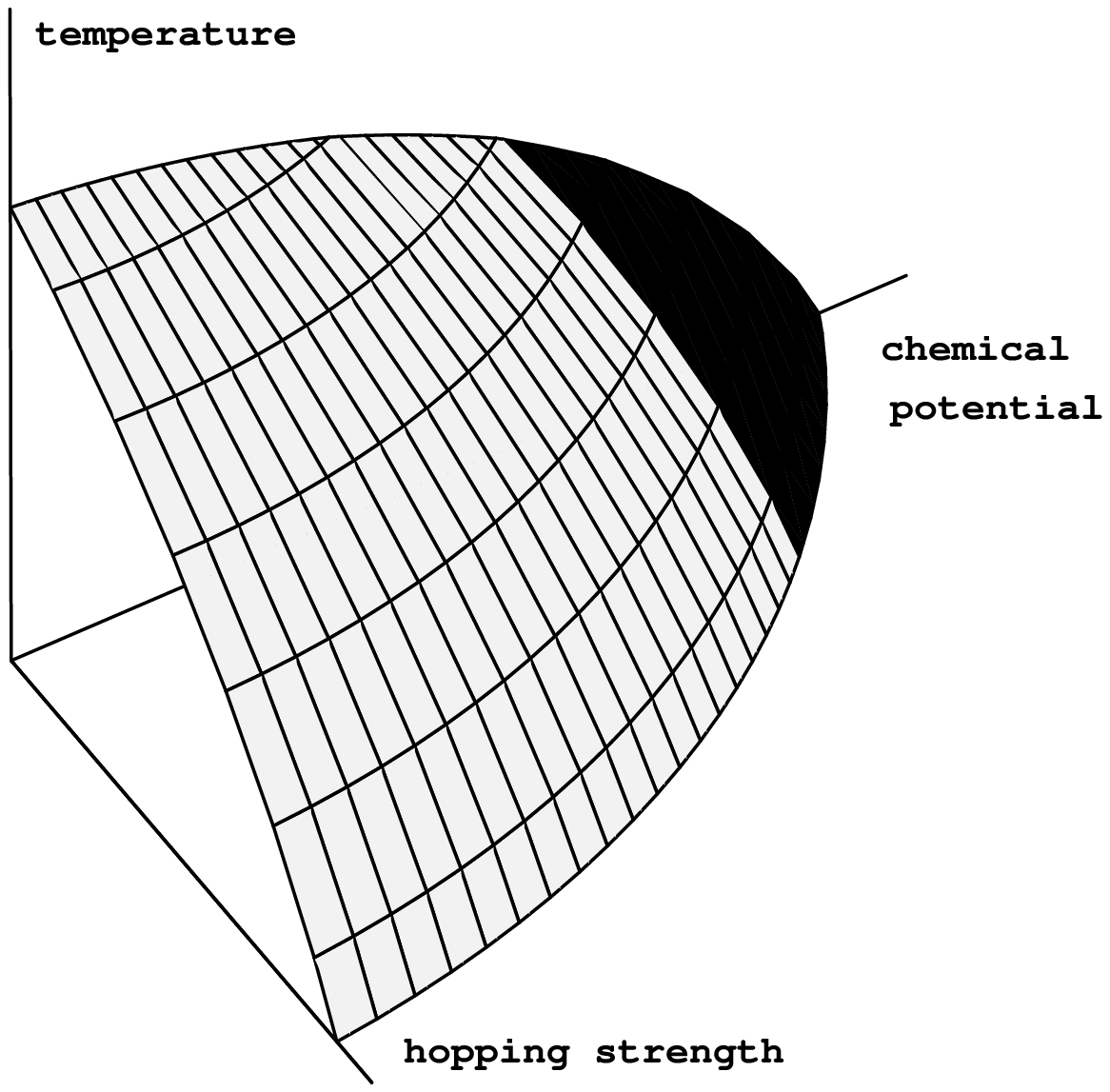,width=7cm,angle=0}
\end{minipage}
\vspace{-7.8cm}
\begin{figure}
\hfill
\begin{minipage}[]{4.7cm}
\caption[]{\it Phase diagram of fermionic spin glasses as function of
chemical potential, hopping, and temperature.
Above the marked area the system is in the paramagnetic, below in the
spin glass phase.
The shaded area corresponds to continuous transition, the black area to
first order transitions. Tricritical points are located at the boundary
of the two regions.}
\end{minipage}
\end{figure}
\vspace{.5cm}

shows surprising similarities
with the thermal tricritical theory, above all the simultaneous occurence
of critical charge and spin fluctuations \cite{SRO}.
However, electron hopping not only lowers the
$T_c$ of 2nd order transitions but also that of tricritical and $1st$
order transitions. For a metallic model with Gaussian random hopping 
and bandwidth $2E_0$
we found a quantum TCP at
%----------------------tricritical quantum transition---------------------
\begin{equation}
E_F = (1-(5/8)^{\frac{1}{2}})^{\frac{1}{2}}E_0 ,\quad\quad  J_c=3\pi
E_0[1-E_F^2/E_0^2]^{^{-\frac{3}{2}}}/32,
\end{equation}
%-------------------------------------------------------------------------
Corrections to this Q - static approximation can be calculated
by generalizing method \cite{MilHu}.
The general phase diagram is sketched in figure 2.
\section{Renormalisation Group Analysis}
\subsection{Tricritical Ising Spin Glass}
We performed a 1--loop RG calculation for 
tricritical fluctuations.
At each RG step the mass of charge fluctuations $\delta Q^{aa}$ was 
shifted away.
Introducing the anomalous dimensions $\tilde{\eta}$ and $\eta$ for diagonal
and offdiagonal fluctuations one finds at one loop level the following
RG relations ($\epsilon = 8 - d$)
%--------------------------RG-flow eqs.------------------------------
\begin{eqnarray}
\frac{dr}{dl}&=&(\frac{d}{2}-11\kappa^2_1+16\kappa_1\kappa_2+6\kappa^2_2)r-
\kappa^2_2,\frac{du}{dl}=2(1-\kappa_1^2)u-4\kappa_1^2+4\kappa_1\kappa_2,
\nonumber\\
\frac{d\kappa_1}{dl}&=&\frac{\epsilon}{2}\kappa_1+9\kappa_1^3,
\frac{d\kappa_2}{dl}=(\frac{\epsilon}{2}+6\kappa^2_2-
\kappa_1^2+16\kappa_1\kappa_2)\kappa_2,
\frac{d\kappa_3}{dl}=(\frac{\epsilon}{2}+9\kappa^2_2)\kappa_3,\nonumber
\end{eqnarray}
Above $d=8$ the RG flows towards the Gaussian fixed point with
mean field exponents, for $d<8$ there is no perturbatively 
accessible fixed point for real $\kappa's$.
However, a preliminary analysis of the resulting strong coupling problem 
shows that
there exists a solution with positive $\tilde{\eta}$ in contrast to the 
negative anomalous dimension typical of cubic field theories with imaginary
coupling.
The RG for the $DIC$ $y_4$ showed that its long--distance behaviour 
is dominated by a $\kappa^4$--contribution (like in \cite{FiSo} but) for 
$d_c^{(u)}=8<d<10$.
This leads to the modified MF exponent 
$\theta_3=8/(d-4)$, which satisfies the scaling relation 
$\theta_3=2/\beta_{\Delta_3}$ in $d_{c3}^{(u)}=8$ 
and reduces to the MF--result 
in $10$ dimensions.
The dimensional shift by $2$ in comparison with \cite{FiSo} is due to 
coupling $t$.

%*************************************************************************

\subsection{Metallic Quantum Spin Glass}

In the theory of quantum phase transitions time--dependent fluctuations
are treated on an equal footing with spatial fluctuations 
\cite{SRO}. While
for the Lagrangian (\ref{seventeen}) it was sufficient to consider only 
the $\omega = 0$ - component of the Q - fields, in the quantum case
all low energy fields must be kept as they are coupled via the quantum
mechanical interaction $u\int d\tau (Q^{aa})^2$. However, the value 
$z = 4 $ of the dynamical
critical exponent renders the u  - coupling dangerously irrelevant
and allows for a perturbative mapping of the critical theory on 
a classical problem, the Pseudo - Yang - Lee edge singularity. 
This field theory has only one cubic coupling which corresponds to
$\kappa_1$ in eq(\ref{seventeen}). The comparison of the metallic 
quantum case with the thermal tricritical theory allows one to understand
the nature of the strong coupling RG fixed point: whereas the spin 
fluctuations in the thermal second order regime are governed by a 
perturbatively accessible fixed point, the TCP and the quantum case
are charcterized by the combination of charge and spin fluctuations 
and a corresponding strong 
coupling problem.

\end{document}